# Additional scaling regions of ion-sputtered surfaces

Oluwole Emmanuel Oyewande

*Department of Physics, University of Ibadan, Ibadan, Nigeria.*

## *ABSTRACT*

*In the continuum theory the time evolution of surfaces eroded by ion bombardment is modelled by stochastic partial differential equations (SPDEs). These determine the scaling regimes and universality classes of the evolving surfaces. Current knowledge of these scaling regimes is based on the existing continuum theory calculations of topographic phase diagrams as functions of the sputtering parameters, which is limited to small values and isotropic cases of the sputtering parameters. And which accounts mainly for ripple patterns and rough surfaces. Recent work has demonstrated the existence of non-ripple nanostructures for anisotropic collision cascade parameters not considered in the existing continuum theory analysis of possible scaling regimes of sputtered surfaces. In this work we calculate phase diagrams representative of a wider range of sputtering conditions, with anisotropic collision cascading in general. The results reveal new scaling regimes yet unaccounted for. The results also reveal a possibility of modelling a wide range of materials with the same SPDE for θ≲30° thus indicating an important universality class.*

***Keywords****: Surface sputter-erosion, nanostructures, nano-pattern formation, stochastic partial differential equations, continuum theory, Monte Carlo simulations, topographic phase diagrams, scaling exponents, universality classes.*

## INTRODUCTION

Surface modification and evolution is a major focus of extensive research on driven non-equilibrium processes occurring on surfaces at nanometre length scales. The evolution can be driven by particle deposition or bombardment. In both cases the surface is described by a continuous and differentiable function $h(x,t)$ which specifies the height of the surface at position $x$ and time $t$, and its evolution by stochastic partial differential equations (SPDEs). The latter process has led to heightened activity in the materials science community in recent years due to the nano-pattern formation accompanying it. It is now well established that under certain sputtering conditions, the random ejection of surface particles due to the arrival of an impinging energetic ion in their vicinity, leads to the formation of ordered nano-structures on the surface. Experiments have demonstrated that the geometry and characteristics of these nanostructures, such as size, roughness, number density, etc., are dependent on the sputtering parameters (Carter & Vishnyakov, 1996; Chason, Mayer, Kellerman, McIlroy, & Howard, 1994; Facsko, Dekorsy, Trappe, Kurz, Vogt, & Hartnagel, 1999; Habenicht, Bolse, Lieb, Reimann, & Geyer, 1999; Carter, 2006; Carter, 2001) which means that these properties of the nanostructures can be controlled by controlling the sputter erosion variables like ion beam incidence angle, penetration depth, and the longitudinal and lateral straggle of the impinging ion within the first few subsurface layers. Thus, ion bombardment induced sputter-erosion processes have great potential for electronic and opto-electronic applications, as well as for nano-device fabrication.

However, due to the complexity of the processes leading to pattern formation, it is often taken for granted that one cannot expect to model the various possible scenarios of different sputtering experiments with the same SPDE or even a simultaneous set of such SPDEs. Moreover the solution of any of the SPDEs that describe a particular experimental setup is difficult due to considerations of an appropriate discretization, analysis of approximation errors and stability. The current approach to a determination of the scaling behaviour of a surface under given experimental conditions is to deduce the SPDE (continuum theoretical model) that characterises the sputtered surface and then to compare it with similar earlier models whose solutions are known (Cuerno & Barabasi, 1995). Things have been made much easier by the work of Cuerno and Barabasi who derived expressions for the coefficients appearing in the continuum model (SPDE) such that one can write a general form for all experiments and then calculate the values of the coefficients for specific experimental sputtering conditions. They calculated phase diagrams for the general case of isotropic and a few cases of anisotropic distribution of the energy of the impinging ion within the material, and reported three scaling regions which are the ones generally assumed to date (Cuerno & Barabasi, 1995; Makeev, Cuerno, & Barabasi, 2002). The existence of only three scaling regions imply that there should be three different sets of scaling exponents of the surface roughness and therefore three universality classes of the sputtered surface. The phase diagram calculations of Cuerno and Barabasi have been limited to sputtering parameters within a smaller region of the phase space than considered in this work; where the two-dimensional phase space is defined in terms of the penetration depth $a$ and the incidence angle $\theta$.

The possibility of the sputtering parameters of some materials falling outside this smaller region has been demonstrated by recent non-atomistic Monte Carlo (MC) simulations (Oyewande, Kree, & Hartmann, 2006) in which atomistic simulations (Ziegler, Biersack, & Littmark, 1985) of the range of ions in solids was performed to determine a realistic range of collision cascade parameters for use in the MC simulations. Moreover, recent work (Oyewande, 2010) has suggested a higher likelihood of this possibility in surfaces created artificially by doping or defect creation. Hence, the need for a more detailed study of a wider range of sputtering conditions. In this paper, we present results of topographic phase diagrams for a wider and more exhaustive range of sputtering parameters. We found new scaling regions yet unaccounted for and provide deeper insights into the complexities of the surface morphology induced by ion bombardment; among which is the result that for incidence angle θ≲30° material surfaces in general fall within the same universality class and can be characterised by the same SPDE, whereas for θ>30° there are several scaling regimes.

**CONTINUUM THEORY**

The continuum theoretical description of interface morphology in terms of deterministic and SPDEs is a powerful and successful tool for understanding the behaviour of diverse interface phenomena. For ion sputtered surfaces, the distribution $E(\boldsymbol{x})$ of the energy $E$ of the incident ion to a surface particle located at position $\boldsymbol{x} = (x_1, x_2, x_3)$ is assumed to be of the Gaussian form (Sigmund, 1969):

$$E(\boldsymbol{x}) = \frac{E}{\left(\sqrt{2\pi}\right)^3 \alpha \rho^2} \exp\left(-\frac{x_3^2}{2\alpha^2} - \frac{x_1^2 + x_2^2}{2\rho^2}\right) \qquad (1)$$

$\alpha$ and $\rho$ are the widths of the distribution parallel and perpendicular to the ion beam direction respectively. The erosion velocity $v \propto \partial_t h$ by definition following which the dynamic evolution of the surface height $h(\boldsymbol{x}, t)$ at nanometre length-scales is for most cases governed by a Kuramoto-Sivashinsky type SPDE (Bradley & Harper, 1988; Cuerno & Barabasi, 1995)

$$\partial_t h(\boldsymbol{x}, t) = -v_0 + \zeta \partial_x h(\boldsymbol{x}, t) + \varsigma_x \partial_{xx} h(\boldsymbol{x}, t) + \varsigma_y \partial_{yy} h(\boldsymbol{x}, t) \qquad (2)$$

$$+\eta_x[\partial_x h(\mathbf{x},t)]^2 + \eta_y[\partial_y h(\mathbf{x},t)]^2 - D\nabla^4 h(\mathbf{x},t) + \beta$$

$v_0$ is the erosion velocity of a flat surface, $\zeta$ is a proportionality constant related to the local surface slope along the $x-$direction, $\varsigma_x$ and $\varsigma_y$ are the (linear) surface tension coefficients, $\eta_x$ and $\eta_y$ are the nonlinear coefficients, $D$ is the (thermal) surface diffusion coefficient, and $\beta$ is a (Gaussian) noise term with zero mean representing the randomness in the ejection of the surface particles.

Cuerno and Barabasi (Cuerno & Barabasi, 1995) derived expressions for the coefficients $\varsigma_x, \varsigma_y, \eta_x$, and $\eta_y$ in terms of the experimental sputtering parameters, these expressions are required for the phase diagram calculations and for ease of reference we provide them as follows. Using the convenient notation:

$$a_\alpha = \frac{a}{\alpha}, a_\rho = \frac{a}{\rho}, \kappa = \cos\theta, \sigma = \sin\theta, \varpi = a_\alpha^2\sigma^2 + a_\rho^2\kappa^2, \Upsilon = \frac{FEPa}{\alpha\rho\sqrt{2\pi\varpi}}\exp\left(-\frac{a_\alpha^2 a_\rho^2 \kappa^2}{2\varpi}\right),$$

where $F$ is the ion flux and $P$ is the proportionality constant between the power deposition and the rate of erosion,

$$\varsigma_x = \Upsilon a \frac{a_\alpha^2}{2\varpi^3}\left(2a_\alpha^4\sigma^4 - a_\alpha^4 a_\rho^2\sigma^2\kappa^2 + a_\alpha^2 a_\rho^2\sigma^2\kappa^2 - a_\rho^4\kappa^4\right), \qquad \varsigma_y = -\Upsilon a \frac{\kappa^2 a_\alpha^2}{2\varpi}$$

$$\eta_x = \Upsilon\frac{\kappa}{2\varpi^4}\Big[a_\alpha^8 a_\rho^2\sigma^4(3+2\kappa^2) + 4a_\alpha^6 a_\rho^4\sigma^2\kappa^4 - a_\alpha^4 a_\rho^6\kappa^4(1+2\sigma^2)\Big]$$
$$- \varpi^2\big[2a_\alpha^4\sigma^2 - a_\alpha^2 a_\rho^2(1+2\sigma^2)\big] - a_\alpha^8 a_\rho^4\sigma^2\kappa^2 - \varpi^4 \qquad (3)$$

$$\eta_y = \Upsilon\frac{\kappa}{2\varpi^2}\left(a_\alpha^4\sigma^2 + a_\alpha^2 a_\rho^2\kappa^2 - a_\alpha^4 a_\rho^2\kappa^2 - \varpi^2\right)$$

There are two interpretations of the implications of the coefficients as weighting factors in the continuum models, one (Cuerno & Barabasi, 1995) considers the nonlinear coefficients to be relevant at higher length-scales than the length-scale of pattern formation, and reconciles the results of experiments performed at different resolutions of the probe instruments (i.e. different length-scales deferring by the order of 10μm) and the other (Oyewande, 2010; Makeev, Cuerno, & Barabasi, 2002) considers the nonlinear effects to be relevant at higher time-scales (of the order of hours) but within the same length-scale and addresses the results of experiments performed for different time durations. In the former, nonlinear effects are irrelevant at the short (nano) length-scales of pattern formation and only become relevant at large length-scales where they destroy the ripple patterns (Cuerno & Barabasi, 1995). Whereas in the latter, nonlinearities may become very relevant at the short (nano) length-scales in which case they may destroy the patterns entirely, leaving a rough surface or reduce the patterns to ordered protrusions (dots) on the surface which exhibit a characteristic length-scale (Oyewande, 2010).

In any case when nonlinearities are irrelevant Eq. 2 predicts the presence of a characteristic length-scale $\Gamma = \sqrt{D/|\varsigma|}$ in the system (Bradley & Harper, 1988; Cuerno & Barabasi, 1995); where $|\varsigma|$ is the largest absolute value of the negative surface tension coefficients. This characteristic length-scale manifests as periodic structures/ripples (e.g. in the separation of ripple crests/troughs) oriented along the direction with the largest $|\varsigma|$. Thus, if neither of $\varsigma_x$ or $\varsigma_y$ is negative the characteristic length-scale is absent and since the ripple wavelength is $\lambda = 2\pi\Gamma\sqrt{2}$, no ripples are formed. When nonlinearities are relevant the first interpretation predicts three scaling regimes characteristic of different evolution rates of the surface roughness, while the second interpretation indicates that the scaling of the surface width is of a more complicated nature (Oyewande, 2010).

**METHODOLOGY**

The phase boundaries are defined by the values of $\theta$ and the dimensionless quantities $a_\alpha, a_\rho$ satisfying the equations

$$\begin{aligned} \nu_x(\theta, a_\alpha, a_\rho) &= \nu_y(\theta, a_\alpha, a_\rho) \\ \lambda_x(\theta, a_\alpha, a_\rho) &= \lambda_y(\theta, a_\alpha, a_\rho) \end{aligned} \quad (4)$$

This is easily done by contour plots of the coefficients with Microsoft Excel chart. However, in order to have a desired contour plot which differentiates between different regions instead of the default contour plotting, the program output was designed to produce results in tabular form such that the row and column labels are, in one instance, the dimensionless quantities $a_\alpha$ and $a_\rho$ while the entries are the labels of the regions (see the Excel worksheet in Figure 1). In this manner we obtained the phase diagrams $a_\alpha$ versus $a_\rho$ for $0 < a_\alpha \leq 6; 0 < a_\rho \leq 6$; $\theta = 50, 70, 75, 85$. We also obtained a phase diagram $\varphi = \alpha/\rho$ against $\theta$. Note that $0 < a_\alpha \leq a_\alpha^m$ is the same as choosing $a = 1.0$, and restricting $\alpha$ to $1/a_\alpha^m \leq \alpha < \infty$.

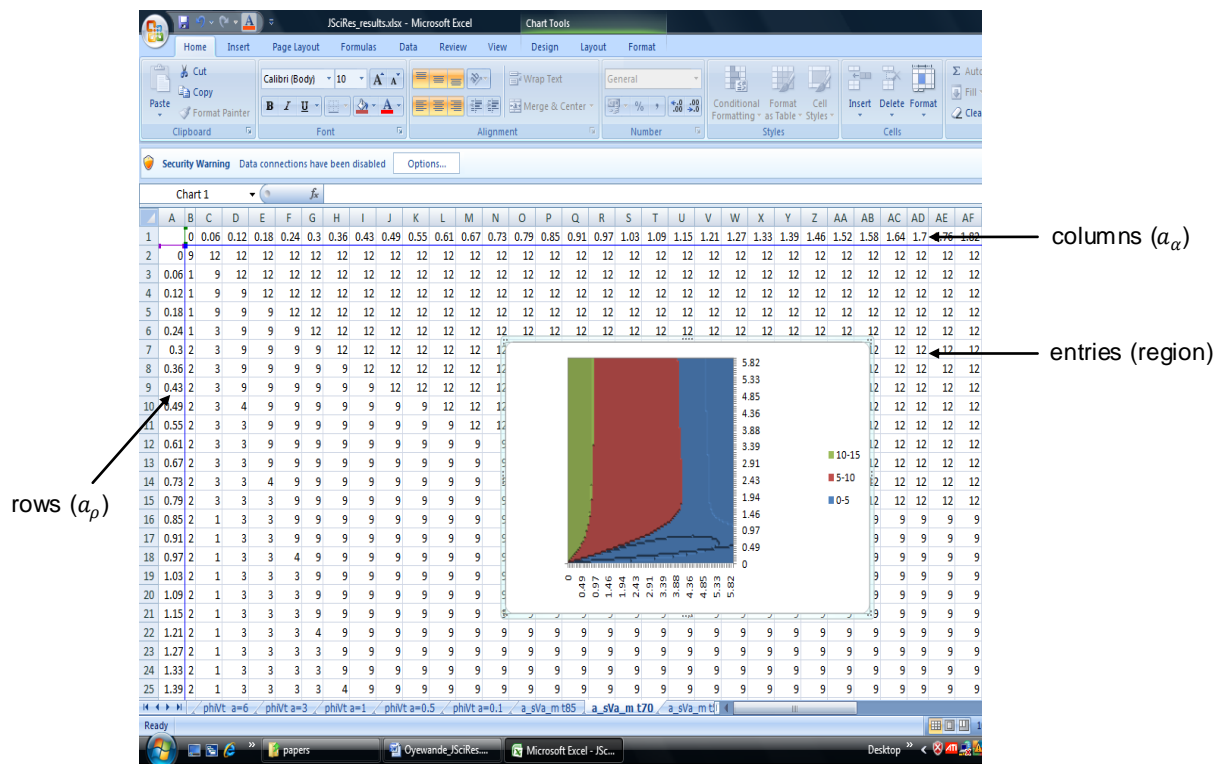

Figure 1: Microsoft excel worksheet showing the output layout for data imported into the excel worksheet from the output data file. The output layout is as in the data file.

Now the coefficients are natural functions of $\theta, a, \alpha,$ and $\rho$ so that the dimensionless coordinates used in Eq. 4 are chosen for their being dimensionless and to reduce the number of independent variables to be monitored in the analysis of the results, since we can fix $\theta$ and consider the coefficients as functions of the size of $\alpha$ and $\rho$ relative to $a$, that is, as functions of the two variables $\alpha/a$ and $\rho/a$ expressed per unit $a$, or $a/\alpha$ and $a/\rho$, instead of the three variables $a, \alpha, \rho$. This means that for a fixed value of $\theta$ all scenarios for which $a \leq \alpha$ are restricted to the region $a_\alpha \leq 1$ in the phase diagram, while all scenarios for which $a$ is greater than $\alpha$ are restricted to the region for which $a_\alpha > 1$ in the phase diagram; and the same argument applies to $\rho$. For a different perspective of the results, we obtained phase diagrams defined in terms of the variables $\varphi = \alpha/\rho$ and $\theta$ to compare the effect of $\theta$ and the relative sizes of $\alpha$ and $\rho$, at fixed values of $a$; meaning that for a fixed value of $a$ all scenarios with $\alpha \leq \rho$ are restricted to the region $\varphi \leq 1$ in the phase diagram and all those with $\alpha > \rho$ are restricted to the region $\varphi > 1$.

## RESULTS AND DISCUSSION

The calculated phase diagrams are presented in Figs. 2 and 3. The colours of the regions in both figures are arbitrary and not specific to particular regions. The regions are defined by the combinations of the coefficients based on their signs or relative magnitudes. Those combinations occurring for the chosen range of parameters studied in this work are defined in Table 1. The currently well-known regions in the literature are the scaling regimes 1, 3, and 9. Figs. 2 and 3 clearly show that the SPDEs, of the form of Eq. 2, which may be used to model sputtered surfaces, are of a more complicated nature as a wide number of possibilities of the relative signs and sizes of the coefficients exist.

| Region | Definition | Region | Definition |
|---|---|---|---|
| 1 | $\varsigma_x < \varsigma_y \leq 0; \eta_x < 0, \eta_y < 0$ | 8 | $\varsigma_y < \varsigma_x \leq 0; \eta_x > 0, \eta_y > 0$ |
| 2 | $\varsigma_y < \varsigma_x \leq 0; \eta_x < 0, \eta_y < 0$ | 9 | $\varsigma_x > 0, \varsigma_y < 0; \eta_x > 0, \eta_y < 0$ |
| 3 | $\varsigma_x < \varsigma_y \leq 0; \eta_x > 0, \eta_y < 0$ | 11 | $\varsigma_x > 0, \varsigma_y < 0; \eta_x > 0, \eta_y > 0$ |
| 4 | $\varsigma_y < \varsigma_x \leq 0; \eta_x > 0, \eta_y < 0$ | 12 | $\varsigma_x > 0, \varsigma_y < 0; \eta_x > 0, \eta_y = 0$ |
| 7 | $\varsigma_x < \varsigma_y \leq 0; \eta_x > 0, \eta_y > 0$ | | |

Table 1: Definition of the possible regions in the range of parameters studied.

Figure 2 indicates the occurrence of the largest number of scaling regimes for $30° \lesssim \theta \lesssim 70°$. Since the quantities $a_\alpha$ and $a_\rho$ characterise different materials (i.e. different widths of the distribution represent different interactions of the ion beam with the surface particles) then Figure 2 reflects the increasing sensitivity of the characteristic SPDE of a material to changes in $\theta$ for values within the above $\theta$ range. The arrows in Figure 2 (d) are indicative of the trend as θ decreases, all the regions except region 2 diminish gradually as their boundaries shift toward the left (direction of the arrows) while region 2 enlarges. For $\theta \approx 30°$ the phase space is almost entirely region 2, which shows the possibility of modelling a wide range of materials with the same SPDE for $\theta \lesssim 30°$. These results (a) – (d) considered for unit $a$ are contained in the result of Figure 3(c) in the second perspective.

Figure 3 confirms the diverse scaling of sputtered surfaces and we find that if $a > 1$ region 2 is more pronounced irrespective of whether $\alpha \lessgtr \rho$. But it should be noted that the results for $a \neq 1$ reflects the effect of a different re-scale of lengths which conveys similar information as the more straightforward result of Figure 3(c). The arrows in Figure 3(d) summarises results for smaller penetration depths $a < 0.5$ where region 1 grows very rapidly to cover almost half of region 2 as $a$ decreases to 0.1 and region 3 grows less rapidly to cover almost the entire region 4. Both figures indicate that regions 2, 9, and 12 are among the most prominent and their corresponding SPDE are therefore among the most relevant universality classes.

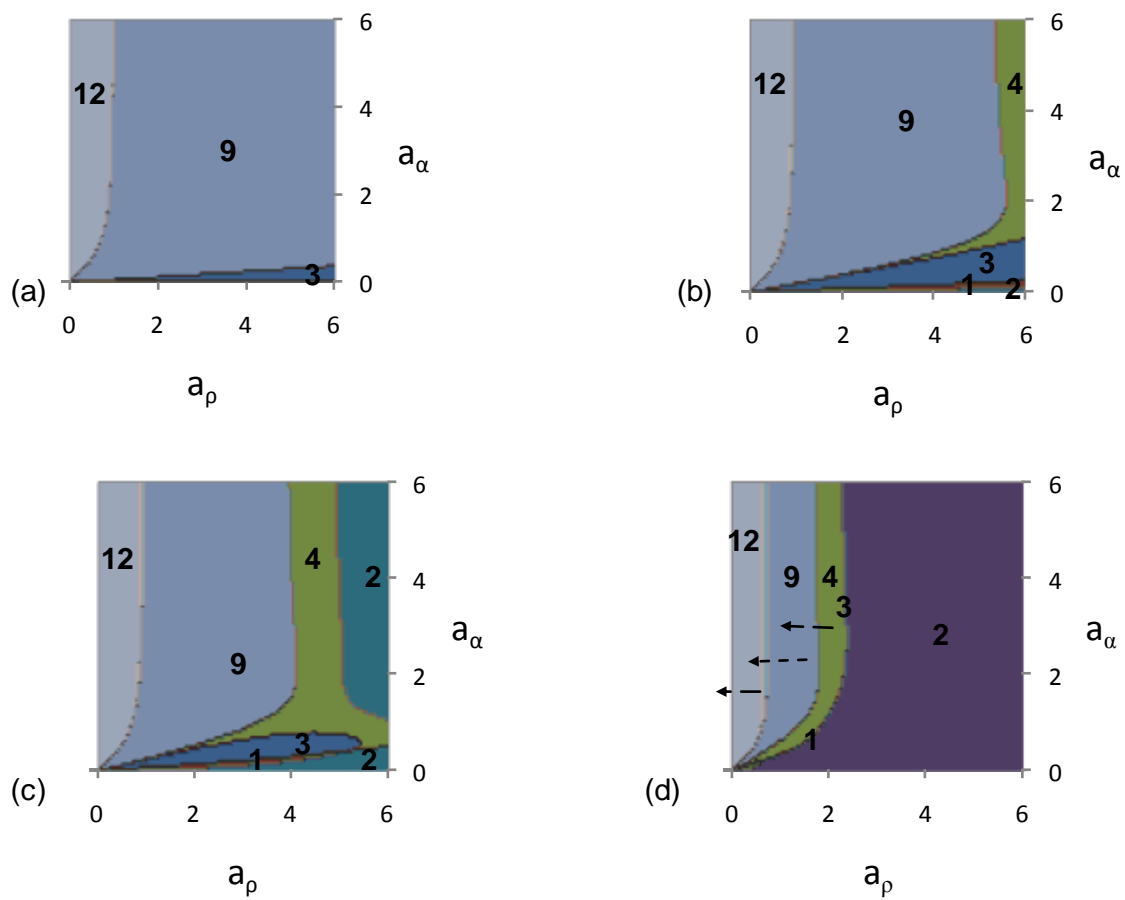

Figure 2: Topographic phase diagrams defined by the dimensionless quantities $a_\alpha = a/\alpha$ and $a_\rho = a/\rho$ for different angle of incidence $\theta$. (a) $\theta = 85$, (b) $\theta = 75$, (c) $\theta = 70$, and (d) $\theta = 50$.

## CONCLUSION AND RECOMMENDATIONS

The current approach to the determination of the scaling behaviour of a surface under given experimental conditions by a comparison of its characteristic SPDE with similar earlier models whose solutions are known, will remain a feasible method if the generality of experimental parameters fall within the small range that can be classified in terms of the three scaling regions currently known. While the results presented here have shown the possibility of most materials exhibiting the same scaling behaviour for θ≲30°, most sputtering experiments exceed θ=30° in which case the results presented here indicate the existence of a large number of possible scaling regimes most of which are yet unaccounted for. Since very little number of solved models exists relative to the diverse variety shown in this work, a large number of possible scaling behaviours remain unresolved and unknown. Hence, the current approach for determining the scaling behaviour of sputtered surfaces is inadequate and a discretization of the continuum equations with error and stability analyses would be a great step forward as this would provide a means for computer simulations of thesurface evolution for any required experimental parameters, thus enabling a determination of the governing scaling laws as well as providing answers to the open questions on the scaling of sputtered surfaces. Ongoing research (Oyewande, 2011) indicates that this discretization will be easier than a numerical integration of the continuum equation.

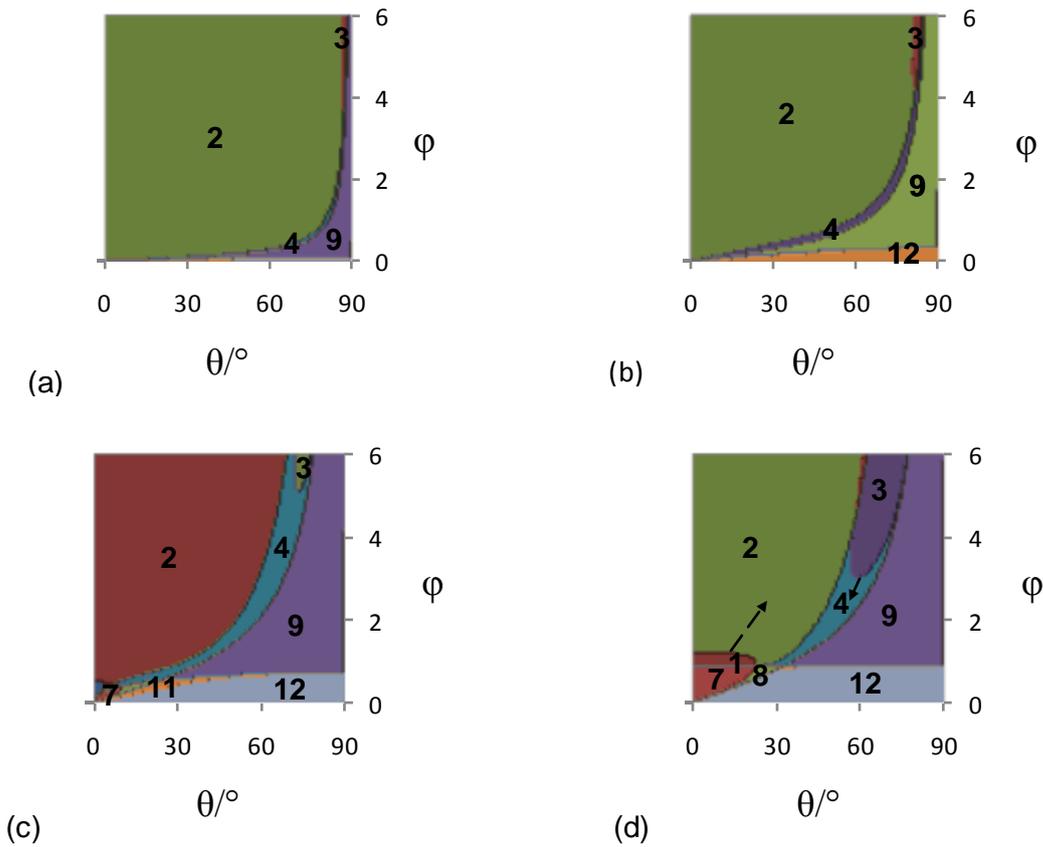

Figure 3: Topographic phase diagrams defined by the dimensionless quantity $\varphi = \alpha/\rho$ and the angle of incidence $\theta$ for different values of the penetration depth $a$. (a) $a = 10$, (b) $a = 3$, (c) $a = 1$, and (d) $a = 0.5$.

An alternative approach to access the scaling laws is the statistical mechanics method of Monte Carlo sampling. Cuerno and Makse introduced a discrete stochastic 1+1 dimensional classical model with assumptions of surface height evolution only by the two competing processes of erosion and surface diffusion. But more experiments are still required to clarify crossovers found between scaling regimes. Furthermore a simple discrete 2+1 dimensional Monte Carlo model was introduced by Hartmann et al. (Hartmann, Kree, Geyer, & Koelbel, 2002) in which details of the atomic dynamics are lost but which affords a statistical sampling of the possible states (i.e. possible height profiles) such that a more realistic probing of the various scaling regimes can be performed by studying the corresponding roughness evolution. Computer simulations with this model (Oyewande, Hartmann, & Kree, 2005; Oyewande, Kree, & Hartmann, 2006; Oyewande, Kree, & Hartmann, 2007) exhibit all the salient features of the height evolution of ion sputtered surfaces but the lack of surface width saturation in several cases of the sputtering parameters frustrates attempts to extract the scaling exponents. Nevertheless this model, due to its robustness and agreement with the published results of the continuum theory is viable and with necessary corrections addressing the surface width saturation could be applied to determine the scaling regimes.

When solving the SPDEs it is recommended that the one representative of region 2 be done first as this is more general after which those of regions 9 and 12 can be done next. In this way the universality class of these important regions can be determined.

e-mail: eoyewande@gmail.com